 \documentclass[prb,twocolumn,showpacs,amsmath,amssymb,superscriptaddress]{revtex4}
 \usepackage{graphicx}
\usepackage{dcolumn}
\usepackage[sort&compress]{natbib}
\usepackage{subfigure}
\usepackage{ifpdf}
\usepackage{bm}
\ifpdf
\usepackage[pdftex,
        colorlinks=true,
        pdftitle={},
        pdfauthor={S. Salem-Sugui Jr.},
        pdfsubject={fluctuation in iron-pnictides},
        pdfkeywords={IF, UFRJ, Rio de Janeiro Brazil},
        baseurl={http://www.if.ufrj.br},
        pdfpagemode=UseNone,
        bookmarksopen=true
        pdfoutput=1
        ]{hyperref}
\fi
\begin{document}
\title{Flux-dynamics associated with the Second Magnetisation Peak in iron-pnictide $Ba_{1-x}K_xFe_2As_2$}
\author{S. Salem-Sugui Jr.}
\affiliation{Instituto de Fisica, Universidade Federal do Rio de Janeiro,
21941-972 Rio de Janeiro, RJ, Brazil}
\author{L. Ghivelder}
\affiliation{Instituto de Fisica, Universidade Federal do Rio de Janeiro,
21941-972 Rio de Janeiro, RJ, Brazil}
\author{A. D. Alvarenga}
\affiliation{Instituto Nacional de Metrologia Normaliza\c{c}\~ao e
Qualidade Industrial, 25250-020 Duque de Caxias, RJ, Brazil.}
\author{L.F. Cohen}
\affiliation{The Blackett Laboratory, Physics Department, Imperial College London, London SW7 2AZ, United Kingdom}
\author{K.A. Yates}
\affiliation{The Blackett Laboratory, Physics Department, Imperial College London, London SW7 2AZ, United Kingdom}
\author{K. Morrison}
\affiliation{The Blackett Laboratory, Physics Department, Imperial College London, London SW7 2AZ, United Kingdom}
\author{J.L. Pimentel Jr.}
\affiliation{Instituto de Fisica, Universidade Federal do Rio Grande do Sul, Porto Alegre, RS}
\author{Huiqian Luo}
\affiliation{National Lab for Superconductivity, Institute of Physics and National Lab for Condensed Matter Physics, P. O. Box 603 Beijing, 100190, P. R. China}
\author{ Zhaosheng Wang}
\affiliation{National Lab for Superconductivity, Institute of Physics and National Lab for Condensed Matter Physics, P. O. Box 603 Beijing, 100190, P. R. China}
\author{Hai-Hu Wen}
\affiliation{National Lab for Superconductivity, Institute of Physics and National Lab for Condensed Matter Physics, P. O. Box 603 Beijing, 100190, P. R. China}
\date{\today}
\begin{abstract}
We report on isofield magnetic relaxation data on a single crystal of $Ba_{1-x}K_xFe_2As_2$ with superconducting transition temperature $T_c$= 32.7 K which exhibit the so called fish-tail effect. A surface map of the superconducting transition temperature shows that the superconducting properties are close to homogeneous across the sample. Magnetic relaxation data, M(t), was used to obtain the activation energy U(M) in order to study different vortex dynamics regimes. Results of this analysis along with time dependent measurements as a function of field and temperature extended to the reversible region of some M(H) curves demonstrate that the irreversibility as well the second magnetization peak position, $H_p(T)$, are time dependent and controlled by plastic motion of the vortex state. In the region delimited by a characteristic field Hon (well below $H_p$), and $H_p$, the vortex dynamics is controlled by collective pinning. For fields below Hon the activation energy, $U_0$, increases with field as expected for collective pinning, but the pinning mechanism is likely to be in the single vortex limit.  
\end{abstract} 
\maketitle 
\section{Introduction}
The recent discovery of the  iron-pnictides superconductor systems \cite{japan,crystal1,crystal} with critical temperatures ranging from 20 to 55 K raised an intense interest on the study of their properties, such as pairing mechanism, themodynamics and transport, normal-state band-structure, etc.. Among these works, one can also find few studies dedicated to the vortex-dynamics, which due to their relatively high  $T_c$  and upper critical field $H_{c2}$, are gaining interest for applications. Iron-pnictides materials, depending on each system and doping, exhibit the peak effect in the critical current, which is associated with a second magnetization peak appearing in the magnetization field M(H) curves. Some of these systems also present a large magnetic relaxation which resembles for instance the giant-magnetic-relaxation observed in the cuprates \cite{yeshurun1}. The study of the second magnetization peak also known as the "fish-tail" peak is of great interest, from both, academic as well technological view points \cite{yeshurun2,lesley1}. Fundamentally speaking, the mechanism and the origin of this effect is still much debated partly because it is system dependent with classification predominantly determined by superconducting anisotropy \cite{yeshurun2,lesley1,lesley2,lesley3,rosenstein1,rosenstein2}. 

So far, flux-dynamics studies of the second magnetization peak in iron-pnictides were performed on the systems, $SmFeAsO_{0.9}F_{0.1}$ with $T_c$= 55 K where the authors infered weak and collective pinning \cite{wen3}, $NdFeAsO_{0.85}$ \cite{moore} and $Ba(Fe_{1-x}Co_x)_2As_2$, the most studied system, where the peak effect appears only for samples near optimally doping \cite{wen2,nakajima,apl,proz,physC,phaset} and weak and collective pinning are claimed in most of the works. The fish-tail has been also observed in $Ba_{0.6}K_{0.4}Fe_2As_2$ with $T_c$=36.5 K and studied from transport measurements \cite{wen1}. It is interesting to mention that the slightly underdoped samples  of the $Ba_{1-x}K_xFe_2As_2$ system ($T_c \gtrsim$30K) presents a phase-separated co-existence of antiferromagnetism and superconductivity \cite{park} which might be associated to the fact that samples with $T_c$ below 28 K do not show the second magnetization peak.  Indeed it is likely that most forms of inhomogeneity ($T_c$ variation, doping variation, impurity phases, magnetic inclusions) will wash out the peak effect. Its observation is usually related to sample purity. Studies on $Ba(Fe_{1-x}Co_x)_2As_2$ includes; determination of the normalized flux-pinning force around the second magnetization peak \cite{apl}; collective to plastic pinning crossover at  the peak suggested by flux-creep data and relaxation rate analysis \cite{proz}; collective to plastic pinning crossover at the peak infered by flux-creep data and the generalized-Inversion-Scheme analysis for the activation energy \cite{wen2}; fish-tail studied by magnetic measurements and magneto-optical imaging \cite{nakajima}; observation of a highly disordered vortex-state from  Bitter decoration and small-angle neutron scattering \cite{physC}; vortex state structural phase transition at the peak from magnetization and flux-creep measurements within a thermodynamics analysis \cite{phaset}. The above studies are summarized in the table below, with comments of the type of measurement performed and the main conclusions of the study.
\begin{table}[t]
\includegraphics[width=\linewidth]{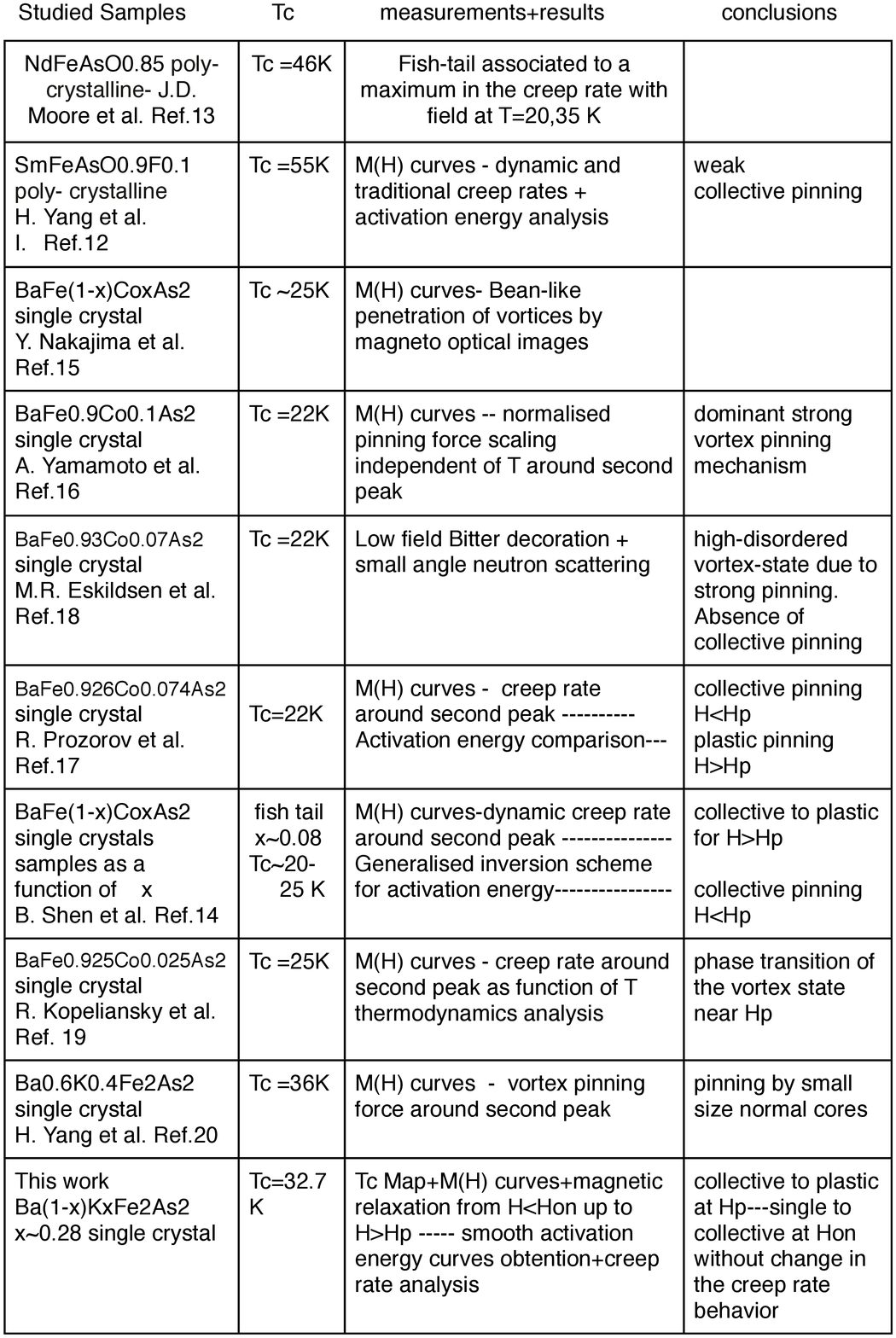}
\caption{Iron-pnictide systems which show the fish-tail effect.}
 \label{tablenew}
\end{table}

Importantly, the possibility that a first order phase transition instead  of a vortex-dynamics crossover has been proposed to explain the second magnetization peak in the pnictides demonstrates the need of more detailed and rigorous  vortex-dynamics analysis. It is that motivated the present work. Another point not covered in the literature is the study of the crossover that should exist at the onset field of the fish-tail, also known as $Hon$. Phenomenologically speaking one should expect a crossover from single to collective pinning at $Hon$ because magnetization changes curvature at $Hon$,  and it appears that this matter has not been studied in detail. 
In this work we address these above points by performing a detailed study of the vortex-dynamics as a function of magnetic field and temperature in an iron pinictide single crystal of $Ba_{1-x}K_xFe_2As_2$ with superconducting transition temperature $T_c$= 32.7 K which exhibits the fish-tail effect.  The work addresses the existence of a change in the pinning mechanism (or crossover) associated with the anomalous second magnetization peak, as found in YBaCuO \cite{abulafia} and  the pinning-crossover expected to exist at $Hon$. This work complements a previous study of the vortex phase diagram performed on the same system \cite{wen1} and as above mentioned vortex-dynamics studies performed in other pinictides systems \cite{wen2,wen3,nakajima, proz} using different approaches. The experiment is conducted by obtaining magnetic relaxation data over selected isothermic M vs. H curves, M(H), for magnetic fields values ranging from just above $H_{c1}$ (actually above the first penetration field peak appearing in isothermic magnetization M(H) curves) up to field values close to the irreversible point Hirr.  Magnetic relaxation curves are used to obtain the corresponding activation energies \cite{maley} allowing the study of the pinning mechanism for magnetic fields in the region of the anomalous second peak as treated in Ref.\onlinecite{abulafia} and below the field $Hon$. We have also measured isofield zero-field-cooled magnetization curves as a function of temperature, $M vs. T$ curves. All data, M(H) and $M vs. T$ curves, were obtained for H$\parallel$ c-axis. The $M vs. T$ curves were used to obtain the near equilibrium irreversibility line, since M(H) curves obtained at different effective $dH/dt$ rates shown that the irreversible point is time dependent.
 
Results of this work show that the temperature dependence of the second magnetization peak position, $H_p$ and of the irreversibility field Hirr are well explained in terms of a plastic motion of the vortex lattice.  Results also demonstrate the existence of a crossover in the pinning mechanism at $Hon$, where apparently,  this crossover occurs without a change in the behavior of the activation energy $U_0$ with field (increasing with field).  
\section{Experimental} 
We measure a high quality crystal of $Ba_{1-x}K_xFe_2As_2$ with $T_c$= 32.7 K corresponding to a potassium content x=0.28 and with mass of approximately 0.05 mg.   This is the same sample studied in our previous work \cite{sugui} and show a fully developed superconducting transition with width $\Delta T_{c}\simeq 1 K$. The crystal was grown by a flux-method described elsewhere \cite{crystal}.  Magnetization and magnetic relaxation data were taken after cooling the sample in zero applied magnetic field (but in the presence of the earth magnetic field). A commercial magnetometer, based on a superconducting quantum interference device (SQUID) was used for bulk magnetization measurements and a scaning Hall probe magnetometer (MHPM)\cite{perkins} was used to map the superconducting transition temperature distribution over the entire sample and to obtain few isothermal images of magnetic field profile in the sample, with a spatial resolution of 5 microns.  Magnetization-vs-field, M(H) curves were obtained at fixed temperatures ranging from 24 to 32 K, for fields up to 50 kOe. All M(H) curves were obtained by extracting the data after the field was stabilized, in most cases the superconducting magnet was set in persistent mode. Additional curves were obtained with the magnet in the non-persistent (driven) mode, in order to obtain the hysteresis curves.  Magnetic relaxation data, M(t) curves, were obtained at $\simeq$60 s intervals over a period of $\simeq$4000 s for fields in the lower branch (increasing field) of selected isothermic hysteresis curves. We also measure long time magnetic relaxation curves over a period of 12 hours for selected values of magnetic field at T = 29.5 K and 28.9 K. Few magnetic relaxation data were measured on the upper branch (decreasing field) of the hysteresis curves to check for data symmetry, which confirm that bulk pinning is dominant for the studied isothermals. We also obtained isofield  zero-field-cooled and field-cooled magnetization curves, $M vs. T$, with fields ranging from 0.05 to 30 kOe. 
\section{Results and discussion}
 \begin{figure}[t]
\includegraphics[width=\linewidth]{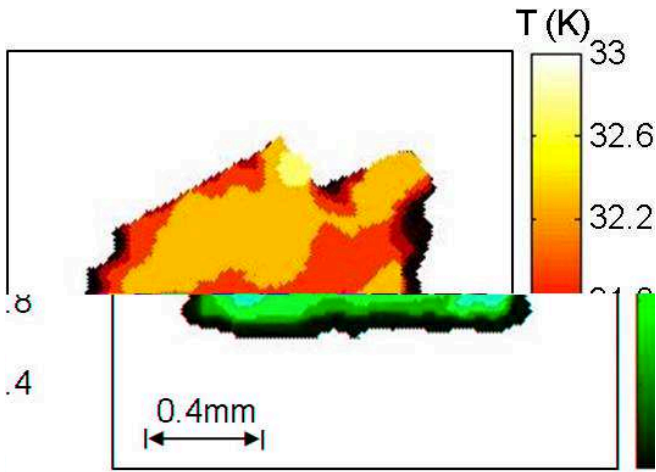}
\caption{Surface map of the superconducting transition temperature of the studied sample. The transition temperatures of the scanned surface are identified by  colors labeled on the right.}
\label{fig1}
\end{figure}  
\begin{figure}[t]
\includegraphics[width=\linewidth]{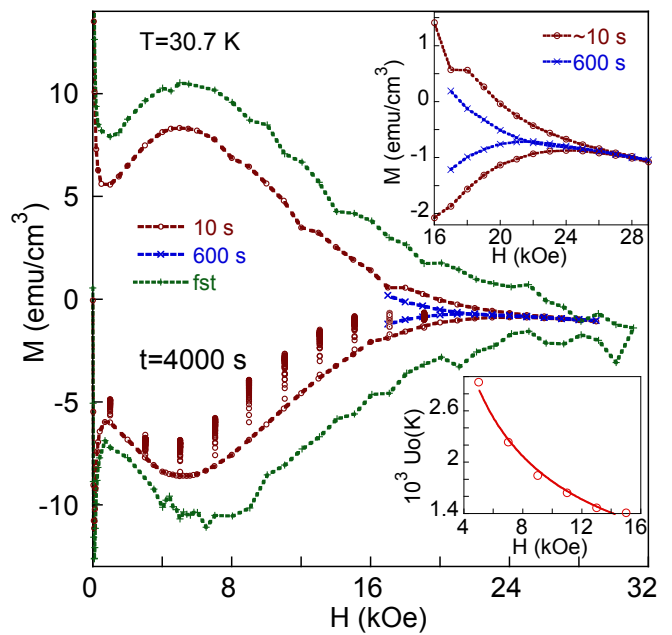}
\caption{Isothermic M(H) curves at T=30.7 K plotted with M(t) data obtained for fixed fields. The upper inset show detail of data for fields near the reversible region. The lower inset show a plot of $U_{0pl}$ vs. H for H$\gtrsim$$H_p$.}
 \label{fig2}
\end{figure}
Figure 1 shows a surface map of the superconducting transition temperature $T_c$ as obtained from the scaning Hall probe magnetometer with a 0.1 kOe field applied parallel to the c-axis after a zero-field-cooled procedure. It is possible to see from Fig. 1 that the sample is basically formed by two major regions, corresponding to more than 90$\%$ of the sample, one with $T_c$=31.8 K  that  surrounds an inner region, the larger one, with $T_c$= 32.3 K. The sample also has a very small border or edge region with $T_c$=31.2 K corresponding to about 5$\%$ of the sample, and an even smaller region with $T_c$ 32.8 K. The $T_c$ homogeneity of the sample can be considered in very good approximation to be within 0.5 K, showing that the sample is of high-quality and any effect due to sample inhomogeneity is expected to be negligible. 
  
Figure 2 shows  typical magnetization curves, M(H), obtained at T=30.7 K exhibiting the anomalous second magnetization peak.  Most of the M(H) curves obtained in this work  show the fish tail as depicted in Fig. 2, allowing the extraction of the values of the fields $Hon$, the field above which the second peak is formed, $H_p$, the field which marks the second peak position, and Hirr which marks the field above which magnetization is reversible. The detail shown by plotting different curves on Fig. 2 is the time dependence of magnetization. The outer curve of Fig.1 was obtained in the fastest way allowed by the equipment  with an effective field sweep rate of $\approx$50 Oe/s, while the internal curve is obtained with an effective field rate of $\approx$1.5 Oe/s. The third curve was obtained near the irreversibility region with a delay time of 10 minutes after each field value is stabilized. The large difference between these three M(H) curves evidences the large magnetic relaxation of this system as reported elsewhere. The upper inset of Fig. 2 shows in detail the relaxation effect on the irreversible field  Hirr. It is possible to see from Fig. 2 that both, $H_p$ and Hirr are time dependent and that their values drop with time. Similar M(H) curves with 10 min time delay as in Fig. 2 were also obtained at T=30.1 and 31.3 K

\begin{figure}[t]
\includegraphics[width=\linewidth]{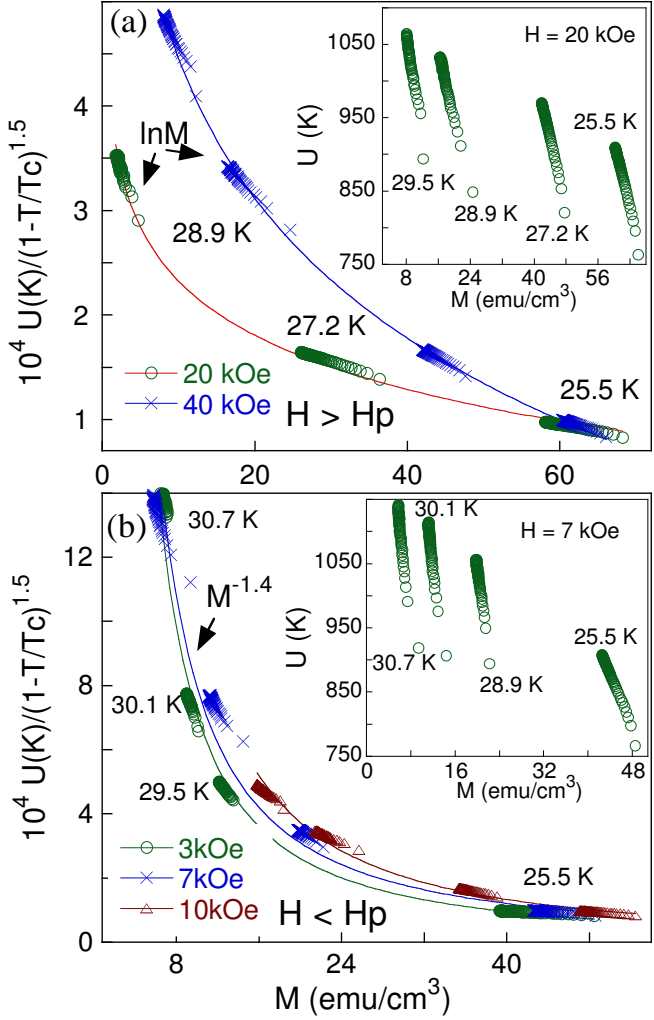}
\caption{Activation energy U(M,T) for fixed fields after scaled by the scaling function $g(T/T_c)=(1-T/T_c)^{1.5}$; a) for fields H$\gtrsim$$H_p$; b) for fields Hon$\lesssim$H$\lesssim$$H_p$. Insets show U(T) vs M curves for selected fields prior scaling}
 \label{fig3}
\end{figure}
Figure 2 also shows magnetic relaxation data measured during 4000 sec for selected fields around the second magnetization peak which are plotted with the original $M(H)$ curve. Magnetic relaxation curves, M(t), as show in Fig. 2 allow to study the vortex-dynamics, and  have been obtained on six isothermic $M(H)$ curves for fields going from below Hon to above $H_p$. Magnetic relaxation curves, M(t),  were collected for a set of selected magnetic fields on MvsH curves at T=25.5, 27.2, 28.9, 29.5, 30.1 and 30.7 K. All M vs. log(t) curves obtained during 4000 seconds follow the typical linear curve observed in most flux creep experiments. This trend was also observed for 12 hours relaxation data obtained for fields above Hon, but not for fields below Hon, as will be discussed later. 

We analyzed flux-creep data by following one of two different approaches. 
Either we obtained the relaxation rate $S=(1/M_0)dM(t)/dlnt$  for each M(t) curve and analyzed its  behavior with H \cite{beasley,donglu,tb}, or  we analyze the activation energy  U(M) by fitting each curve to the expression predicted by the collective pinning theory,  $U(H,M)=U_0(H)(M(t)/M_0)^{\nu}$, which allow to study the behavior of the exponent $\nu$ with H. In the last expression M(t) replaces M(t)-$M_{eq}$ where $M_{eq}$ is the equilibrium magnetization, obtained from the average magnetization of both branches in each M(H) curve.
 
Several different approaches presented in the literature \cite{maley,suvankar,griessen,haihu} allow one to obtain the activation energy U(M) from M(t) curves. Here,  the activation energy U(M) is  obtained for each M(t) curve by applying an approach developed by Maley et al. \cite{maley} where $$U=-Tln(dM(t)/dt)+CT$$ and C is a constant which depends on the hoping distance of the vortex, the attempt frequency and the sample size.  We should mention that similar U(M) curves can also be obtained from M(t) curves  by following an approach developed in Ref.\onlinecite{suvankar}. The insets of Figures 3a and 3b show results of this approach with C=27 (this value will be justified below)  after application to selected M(t) curves obtained for fixed magnetic fields at selected temperatures. The temperatures are selected in a way that each M(t) curve for a given field is located below the field $H_p$ of its original M(H) curve (but above Hon) as in Fig. 3a   and above $H_p$ as in Fig. 3b . This condition is necessary since pinning mechanisms below and above $H_p$ might be of different nature \cite{abulafia}. 
As shown in these figures, the U(M) curves do not form a smooth curve with M, which is expected for temperatures very close to $T_c$ \cite{maley2}. To obtain a smooth curve we have to scale the activation energy curves shown in the insets of Figs. 3a and 3b by a $g(T/T_c)$ scaling function. 

Figures 3a and 3b show the results obtained by choosing $g(T/T_c)=(1-T/T_c)^{1.5}$. This scaling function of U(M) was suggested in  Ref.\onlinecite{maley2} and relies on pinning length scales for temperatures close to $T_c$ \cite{tinkham}. The interesting result of Fig. 3 is that below $H_p$ (Hon$\lesssim$H$\lesssim H_p$), the smooth curves follow a power law with $M^{-1.4}$ as expect from the collective pinning theory \cite{abulafia}, but a logM behavior is obtained for fields above $H_p$. The above analysis yield a constant C=27 for our sample which is used to obtain the activation energy U(M) for each M(t) curve (we mention that this value of C=27 is of the same order as values obtained for high-$T_c$ cuprates \cite{maley2}).

\begin{figure}[t]
\includegraphics[width=\linewidth]{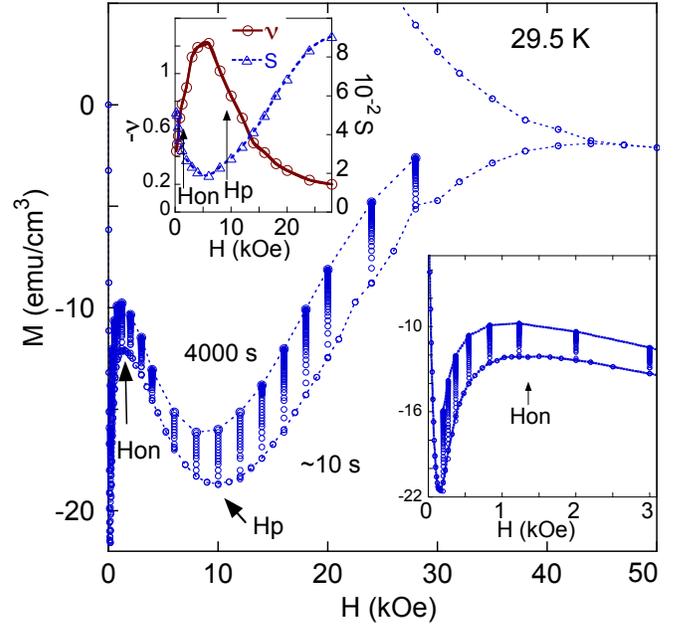}
\caption{Isothermic M(H) curves at  T=29.5 K plotted with M(t) data obtained for fixed fields. The lower inset show detail of data for low fields. The upper inset show a double plot of the exponent $\nu$ vs. H (left y-axis) and S=(1/$M_0$)dM/dlnt vs. H (right y-axis).}
 \label{fig4new}
\end{figure}
Figure 4  shows a selected M(H) curve measured at T=29.5 K. The M(H) curve is plotted with magnetic relaxation data obtained over 1 hour (4000 sec)  for fixed magnetic fields going from below $Hon$ to above $H_p$.  The lower inset of Fig. 4 shows details of the low field data, providing evidence of the behavior before and after the field Hon. The upper inset show results of the relaxation rate S (right y- axis) and exponent $\nu$ (left y-axis) plotted as a function of H. The values of S and of the exponent $\nu$ were obtained as discussed above, by analyzing each M(t) curve and the respective U(M) curve.  It is interesting to note that the behavior of S and $\nu$ with H are quite similar. The relaxation rate drops as field increases from below Hon up to a field close to $H_p$ (the second peak position), increasing again as H become larger than $H_p$. Similar curves for S and $\nu$ as a function of H were observed on all M(H) curves over which we measured magnetic relaxation. 

In addition to the fact that the exponent $\nu$ follows the inverse trend of the relaxation rate S as a function of H, the absolute values of $\nu$ can provide relevant information about the pinning mechanism \cite{abulafia}. As shown in the upper inset of Fig. 4, the region of fields between Hon and $H_p$ corresponds to the region where $-\nu \approx 1$ for which the relation $U(H,M)=U_{0col}(H)(M(t)/M_0)^{\nu}$ predicted by the collective pinning theory might apply. Values of $-\nu \approx 1$ are expected from collective pinning theory, while values much smaller  than 1, as observed on M(t) curves below Hon and above $H_p$, may be due to single vortex pinning regime for lower fields, or plastic pinning occurring above $H_p$ , respectively \cite{abulafia}.  

Figures  5a and 5b show U(M) curves as obtained from M(t) curves appearing in Fig. 4. Figure 5a show a set of U(M) curves for fields H$\lesssim$Hon plotted with a set of U(M) for Hon$\lesssim$H$\lesssim$$H_p$. Figure 5b shows a set of U(M) curves for fields Hon$\lesssim$H$\lesssim$$H_p$ plotted with a set of U(M) curves for H$\gtrsim$$H_p$. It is possible to see from these plots that the fields Hon and $H_p$ are in fact characteristic fields separating regions with  differences in the vortex dynamics. 
Following the results of Ref.\onlinecite{abulafia} we fit U(M) for fields above $H_p$ with the expression $U_{pl}=U_{0pl}(H)(1-\sqrt{M(t)/M_{0pl}}$. This is the appropriated description for plastic motion.  Results of the fitting of U(M) to the collective pinning expression of U performed for Hon$\lesssim$H$\lesssim$$H_p$ and to the plastic expression of U performed for fields H$\gtrsim$$H_p$ produced values of $U_{0col}(H)$ and $U_{0pl}(H)$ which show a power law behavior with H. As expected, $U_{0col}(H)$$\approx$$H^{0.4}$ increases with field while $U_{0pl}(H)$$\approx$$H^{-0.7}$ decreases with field. The above exponents of H are used to scale the correspondents U(H,M) curves shown in Figures 5a and 5b turning them in-to smooth curves of U(M) (in arbitrary units) plotted against M(t). The results of this scaling are shown in Figure 5c.  It is interesting to observe that each scaled curve follows a power law behavior with M where each exponent value agrees with the averaged value of the exponent $\nu$ found in each corresponding field region as shown in the upper inset of Fig. 4.  The smooth curves of Fig. 5c demonstrate the existence of a crossover in the pinning mechanism as field increases above $Hon$ as well demonstrate that the second magnetization peak in the studied sample is due to a pinning crossover mechanism, as was first demonstrated in Ref.\onlinecite{abulafia} for $YBaCuO$. The second magnetization peak occurring at $H_p$ is formed by a crossover in the pinning mechanism, from collective to plastic pinning as the field increases.
\begin{figure}[t]
 \includegraphics[width=\linewidth]{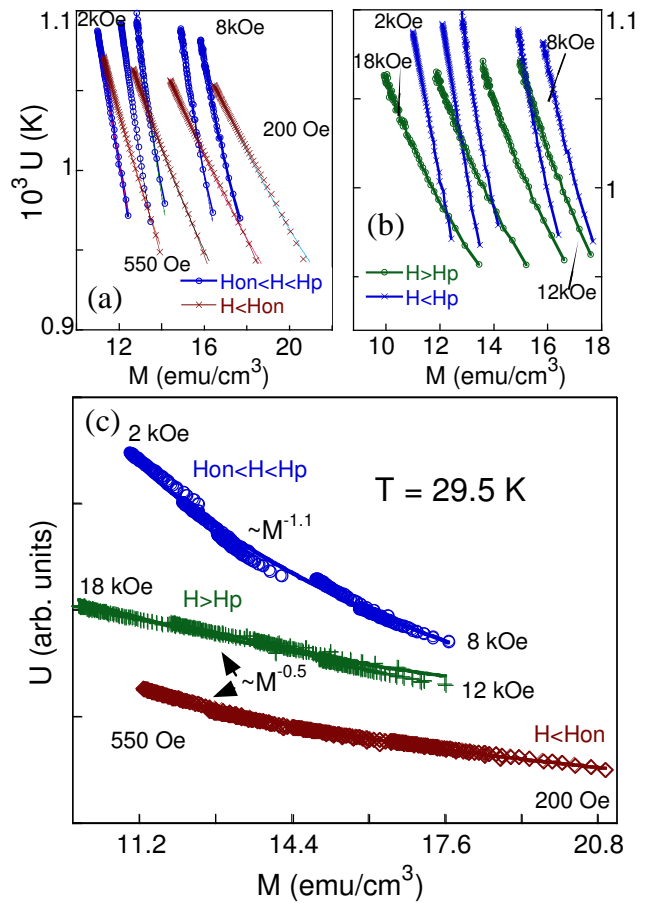}
\caption{U(M) curves as obtained from M(t) curves at 29.5 K: a) U(M) curves for fields H$\lesssim$Hon and Hon$\lesssim$H$\lesssim$$H_p$; b) U(M) curves for fields Hon$\lesssim$H$\lesssim$$H_p$ and H$\gtrsim$$H_p$; c) U(M) curves after scaled.}
 \label{fig5}
\end{figure}
A visual inspection of U(M) curves for H$\lesssim$Hon in Fig. 5a suggests that these curves have the same behavior as the curves obtained for H$\gtrsim$$Hp$ for which it is possible to infer that a plastic pinning dominates. However, this hypothesis is inconsistent with the fact that the activation energy $U_{0pl}(H)$ (found by fitting U(M) curves for H$\lesssim$Hon to the correspondent expression for plastic pinning) for H$\lesssim$Hon increases with field. On the other hand, the scaled U(M) function appearing in Fig. 5c for H$\lesssim$Hon was obtained assuming that  U(M) has a power law dependence with H of the form $\approx H^{-0.2}$, an H dependence with a negative exponent  as found in the region H$\gtrsim$$Hp$. These contradictory facts eliminate the possibility of plastic pinning in the region below Hon, as well as eliminate the possibility of collective pinning as observed for Hon$\lesssim$H$\lesssim$$H_p$. Figure 6 show plots of the relaxation rate $S=(1/M_0)dM/dlnt$ as obtained from magnetic relaxation data over three selected isothermal M(H) curves for fields in the region of Hon. The same trend shown in Fig. 6 of S decreasing with field was observed in all M(H) isothermals,Ê which means that, in fact, the correspondent activation energy $U_0=k_BT/S$ as defined by Beasley et al.\cite{beasley} increases with field in the region $H\lesssim Hon$. This is an interesting finding, because due to the positive inclination of M(H) for H$\lesssim$Hon one would expect $U_0$ to decrease with field. Since above Hon, the activation energy $U_{0col}(H)$ (collective pinning region) also increases with field, the change in the pinning mechanism occurring at Hon has a different nature (probably single to collective pinning)  than the pinning crossover occurring near $H_p$ (which is collective to plastic).   The formation of the peak appears then to come from the existence of a pinning crossover, probably from single to collective pinning, as Hon is crossed. This is the field region (H$\lesssim$Hon) above which the fish-tail shape takes place. 
\begin{figure}[t]
\includegraphics[width=\linewidth]{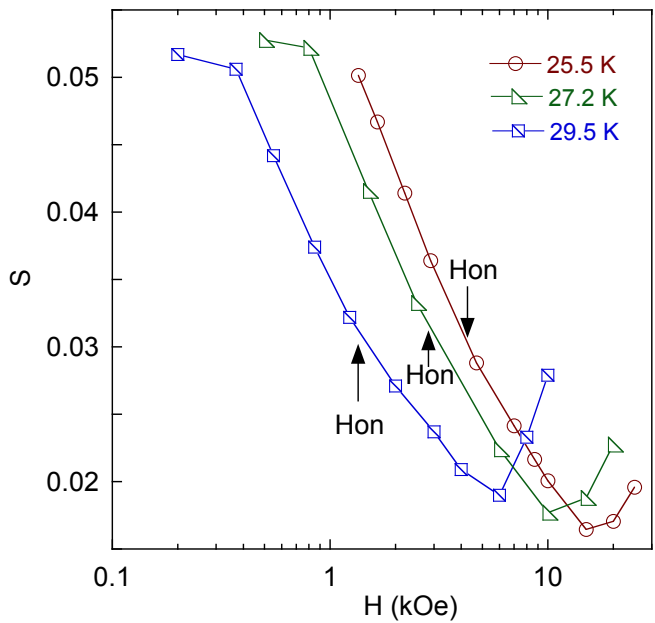}
\caption{The relaxation rate S is plotted against the magnetic field in logarithmic scale for selected temperatures. The plot displays values of S as obtained for fields below Hon up to fields close to $H_p$.}
 \label{fig6new}
\end{figure}
 
\begin{figure}[t]
\includegraphics[width=\linewidth]{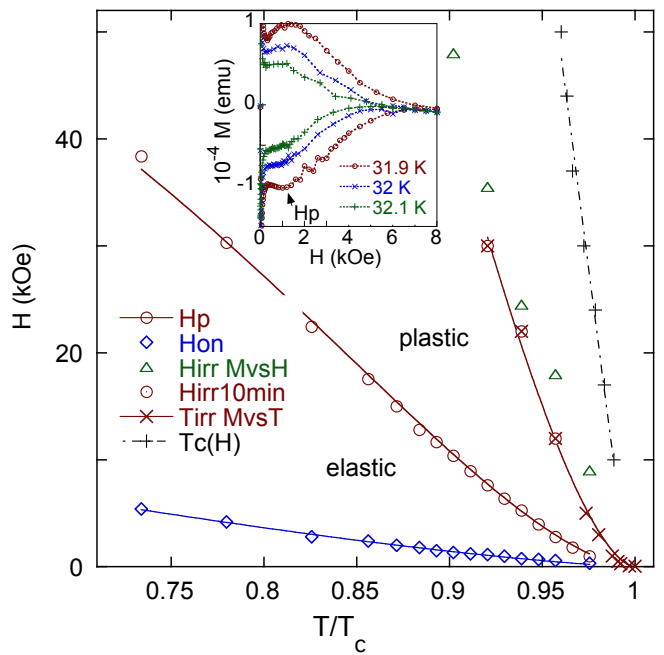}
\caption{Vortex phase diagram of the studied sample. Solid lines on Tirr(T) and $H_p(T)$ data were obtained by fitting the respective data to the expression $\approx [1-(T/T_c)^4]^{1.7}$ , solid line on Hon(T) data was obtained by fitting the data to the expression  $\approx (1-T/T_c)^{1.35}$, while dotted line on $T_c(H)$ data is only a guide to the eyes. The inset show M(H) curves obtained near $T_c$ detailing the temperature onset of the anomalous second magnetization peak.}
 \label{fig7new}
\end{figure}

We present in Fig. 7 the vortex-phase diagram obtained from the M(H) curves. An interesting finding is that the line defined by the values of $H_p(T)$ does not touch the irreversibility line, but ends at some temperature below $T_c$. This feature is shown in the inset of Fig. 7 which shows that the anomalous second peak in the magnetization is only well defined for temperatures below 32 K. A similar behavior for the $H_p(T)$ line was observed for a deoxygenated YBaCuO crystal \cite{ghiv}. The crossover from collective to plastic pinning only exists below
T$\lesssim$32 K, since M(H) curves obtained above this temperature do not show the second magnetization peak. 
It is important to mention again that both, $H_p(T)$ as well Hirr(T), are time dependent, and U(M) curves for fields close to both, $H_p(T)$ and Hirr(T), seems to be fitted by the plastic expression for the activation energy. To exemplify the time dependence of Hirr, values of Hirr obtained on M(H) curves by using different time windows, as for instance show in the upper inset of Fig. 2, are plotted with values of $T_{irr}$ obtained from isofield M vs. T curves (not shown). It is interesting to note that the values of $T_{irr}$ are close to the values of Hirr obtained 10 minutes after the field was stabilized.  As shown in Ref.\onlinecite{abulafia} , the fact that both, $H_p(T)$ and Hirr(T) are time dependent (with these values being shifted to the left with time) suggests that these fields are controlled by plastic pinning with expected temperature dependence of the form  $H_p(T)$$\approx$$[1-(T/T_c)^4]^{1.4}$. This expression was obtained in Ref.\onlinecite{abulafia} after considering that $U_{0pl} \approx H^{-0.7}$. Since our fittings of U(M) in the region H$\gtrsim$$H_p$ produced a similar behavior for $U_{0pl}$ with H, with an exponent of H varying from -0.6 to -0.7 depending on the temperature of the M(H) curve, we also start our fittings of the $H_p(T)$ and Tirr(H) lines by assuming the temperature dependence $\approx$$[1-(T/T_c)^4]^{1.4}$. However, the best fittings, as shown in Fig. 7, were obtained with a slightly different expression, $\approx [1-(T/T_c)^4]^{1.7}$ . For consistence we only fit values of Tirr(H). For the Hon(T) line we try the general expression $\approx (1-T/T_c)^m$ where $m$ is a fitting parameter. (This expression failed to fit Hirr(T) data as well as the high temperature region of $H_p(T)$ data). The best fit  show in Fig. 7 was obtained with $m$=1.35. This value of the exponent $m$ is very close to 1.5, and we observe that the value of $m$=1.5 (corresponding to the temperature dependence of the pinning length scale assumed for $g(T/T_c)$) also produced a reasonable fitting of Hon(T) data suggesting single pinning of vortices in the region below Hon. The dotted line linking the $T_c(H)$ points is only a guide to the eyes. The values of $T_c(H)$ plotted in Fig. 7 were extracted from results obtained in the same sample in Ref.\onlinecite{sugui}. 

\begin{figure}[t]
\includegraphics[width=\linewidth]{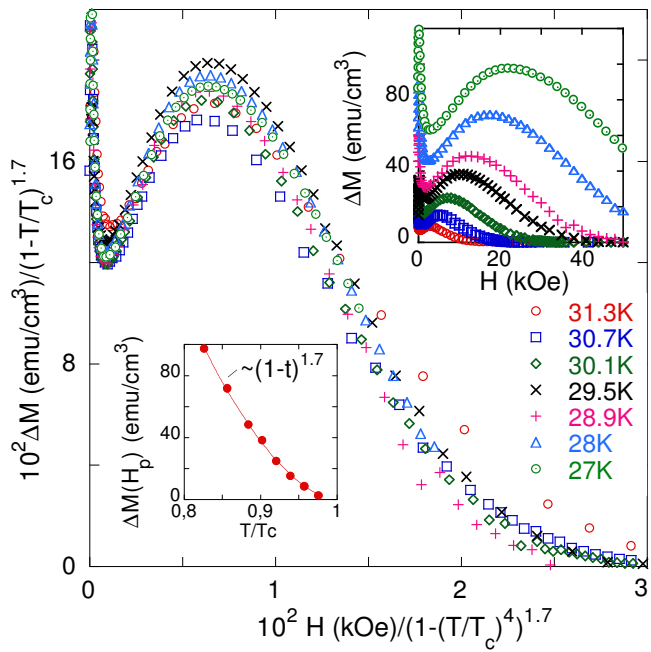}
\caption{Isothermic $\Delta M$/$(1-(T/T_c)^4)^{1.7}$ are plotted against H/$(1-T/T_c)^{1.7}$. Upper inset shows  isothermic  $\Delta M(H)$ curves used in the main figure. The lower inset show a plot with values of $\Delta M$ extracted from curves in the upper inset for H=$H_p$ plotted against the reduced temperature $T/T_c$.}
\label{fig8}
\end{figure}  

We plot in Fig. 8 an attempt to scale several $\Delta M$ vs. H curves which were obtained by subtracting the two branches of each respective M(H) curve. The main figure shows a plot of several $\Delta M$/$(1-(T/T_c)^4)^{1.7}$  vs. H/$(1-T/T_c)^{1.7}$ curves which reflects the consistency of the analysis performed in this work. The upper inset shows the original $\Delta M$ vs. H curves. Since the $H_p$ vs. T line in Fig. 7 follows a dependence with $(1-(T/T_c)^4)^{1.7}$, we find it natural to choose the same temperature dependence to scale the x-axis which is the magnetic field H.  The scaling law used in the y-axis, $\Delta M$, was based on the fact that the strength of the critical current which is of the order of $\Delta M$, should follow the temperature dependence of the pinning length scale, which for temperatures close to $T_c$ has the form\cite{tinkham}  $(1-T/T_c)^{1.5}$ (this is the same scaling function $g(T/T_c)$ used on the analysis of the activation energy presented in Fig. 2). The lower inset of Fig. 8 shows a plot of $\Delta M$ for H=$H_p$ vs. $T/T_c$, which shows a dependence with $(1-T/T_c)^{1.7}$ instead $(1-T/T_c)^{1.5}$. For this reason we choose to scale $\delta M$ with $(1-T/T_c)^{1.7}$ instead $(1-T/T_c)^{1.5}$ which produced a better scaling of the curves. 

\section{Conclusions}
In conclusion, our study of the vortex-dynamics in  $Ba_{1-x}K_xFe_2As_2$ shows that the second magnetization peak occurring at $H_p$  is formed by a crossover in the pinning mechanism, from collective to plastic pinning as the field increases. This crossover only exists below a certain temperature T$\lesssim$32 K, since M(H) curves obtained above this temperature do not show the second magnetization peak. It is also shown the existence of a pinning crossover, probably from single to collective pinning, as  $Hon$ is crossed. This is the field region (H$\lesssim$Hon) above which the fish-tail shape takes place. Results of this work show that both $H_p(T)$ as well Hirr(T) are time dependent and their temperature dependence are well explained by an expression predicted by assuming a plastic motion of the vortex state. We also show that the $g(T/T_c)$ scaling function of U(M) curves (used in Fig. 3) and the temperature dependence expression used to fit the $H_p(T)$ and Hirr(T) lines, can be used to scale several $\Delta M(H)$ producing a reasonable collapse of the curves.

SSS, LG and ADA thanks support from the Brazilian agencies CNPq and FAPERJ. KM,KY and LFC thank the UK Funding Council the EPSRC grant EP/H040048. We thank Y. Yeshurun for a helpful discussion.


\begin{thebibliography}{99}
\bibitem{japan}Y. Kamihara, T. Watanabe, M. Hirano, and H. Hosono, J. Am. Chem. Soc. {\bf 130}, 3296 (2008).
\bibitem{crystal1}M. Rotter, M. Tegel, and D. Johrendt, Phys. Rev. Let {\bf 101} 107006 (2008). 
\bibitem{crystal}H.Q. Luo, Z.S. Wang, H. Yang, P. Cheng, X. Zu, and Hai-Hu Wen, Supercond. Sci. Technol. {\bf 21}, 125014 (2008).
\bibitem{yeshurun2}Y. Yeshurun and A. P. Malozemoff, and A. Shaulov, Rev. Mod. Phys. {\bf 68}, 911 (1996).
\bibitem{lesley1}L.F. Cohen, G. Perkins, J. Laverty, W. Assmus and A.D. Caplin, Cryogenics {\bf 33}, 356 (1993).
\bibitem{lesley2}L.F. Cohen, H. Jensen, Reports on Progress in Physics {\bf 60}, 1581 (1997).
\bibitem{rosenstein2}B. Rosenstein, B.Ya. Shapiro, I. Shapiro, Y. Bruckental, A. Shaulov, and Y. Yeshurun, Phys. Rev.B {\bf 72}, 144512 (2005). 
\bibitem{yeshurun1}Y. Yeshurun and A. P. Malozemoff, Phys. Rev. Lett. {\bf 60}, 2202 (1988).
\bibitem{rosenstein1}B. Rosenstein and V. Zhuravlev, Phys. Rev. B {\bf 76} (2007). 
\bibitem{lesley3}G. Perkins, L.F. Cohen, A.A. Zhukov and A.D. Caplin, Phys. Rev. B {\bf 51}. 8513 (1995).
\bibitem{abulafia}Y. Abulafia, A. Shaulov, Y. Wolfus, R. Prozorov, L. Burlachkov, Y. Yeshurun, D. Majer , E. Zeldov, H. W�hl, V. B. Geshkenbein, and V. M. Vinokur, Phys. Rev. Lett {\bf 77}, 1596 (1996). 014507 (2007).
\bibitem{wen3}H. Yang, C. Ren, L. Shan, and Hai-Hu Wen, Phys. Rev. {\bf 78}, 092504 (2008).
\bibitem{moore}J.D. Moore, L.F. Cohen, Y. Yeshurun, A.D. Caplin, K. Morrison, K.A. Yates, C.M. McGilvery, J.M. Perkins, D.W. McComb, C. Trautmann, Z.A. Ren, J. Yang, W. Lu, X.L. Dong, and Z.X. Zhao, Sup. Sci. Techn. 
\bibitem{wen2}B. Shen, P. Cheng, Z. Wang, L. Fang, C. Ren, L. Shan, and H.-H. Wen, Phys. Rev. B {\bf 81}, 014503 (2010).
\bibitem{nakajima}Y. Nakajima, T. Taen, and T. Tamegai, J. Phys. Soc. Jpn. {\bf 78} 023702 (2009)
\bibitem{apl}A. Yamamoto, J. Jaroszynski, C. Tarantini, L. Balicas, J. Jiang, A. Gurevich, D.C. Larbalestier, R. Jin, A.S. Sefat, M.A. McGuire, B.C. Sales, D.K. Christen, and D. Mandrus, Appl. Phys. Lett. {\bf 94}, 062511 (2009).
\bibitem{proz}R. Prozorov, N. Ni, M. A. Tanatar, V. G. Kogan, R. T. Gordon, C. Martin, E. C. Blomberg, P. Prommapan, J. Q. Yan, S. L. Bud'ko, and P. C. Canfield, Phys. Rev. B  {\bf 78}, 224506 (2008).
\bibitem{physC}M.R. Eskildsen, L.Ya. Vinnikov, I.S. Veshchunov, T.M. Artemova, T.D. Blasius, J.M. Densmore , C.D. Dewhurst, N. Ni, A. Kreyssig, S.L. BudÕko, P.C. Canfield, A.I. Goldman, Phys. C {\bf 469}, 529 (2009).
\bibitem{phaset}R. Kopeliansky, A. Shaulov, B. Ya. Shapiro, and Y. Yeshurun, B. Rosenstein, J.J. Tu, L.J. Li, G.H. Cao, and Z.A. Xu, Phys. Rev. B {\bf 81}, 092504 (2010). 
\bibitem{wen1}H. Yang, H.Q. Luo, Z.S. Wang, and Hai-Hu Wen, Appl. Phys. Lett. {\bf 93}, 142506 (2008).
\bibitem{park}J.T. Park, D.S. Inosov, Ch. Niedermayer, G.L. Sun, D. Haug, N.B. Christensen, R. Dinnebier, A.V. Boris, A.J. Drew, L. Schulz, T. Shapoval, U. Wolff, V. Neu, Xiaoping Yang, C.T. Lin, B. Keimer, and V. Hinkov, Phys. Rev. Lett {\bf 102}, 117006 (2009).
\bibitem{maley}M. P. Maley, J. O. Willis, H. Lessure and M. E. McHenry, Phys. Rev. B {\bf 42}, 2639 (1990).
\bibitem{suvankar}S. Sengupta, D. Shi, S. Salem-Sugui, Jr., Z. Wang, P.J. McGinn, and K. DeMoranville, J. Appl. Phys. {\bf 72}, 592 (1992).
\bibitem{griessen}R. Griessen, Wen Hai-hu, A. J. J. van Dalen, B. Dam, J. Rector, and H. G. Schnack, S. Libbrecht, E. Osquiguil, and Y. Bruynseraedem Phys. Rev. Lett {\bf 72}, 1910, (1994).
\bibitem{haihu}H. H. Wen, H. G. Schnack, R. Griessen, B. Dam and J. Rector, Physica C {\bf 241}, 353 (1995). 
\bibitem{maley2} M. E. McHenry, S. Simizu, H. Lessure, M. P. Maley and J. Y. Coulter, I. Tanaka and H. Kojima, Phys. Rev. B {\bf 44}, 7614 (1991). 
\bibitem{sugui}S. Salem-Sugui, Jr., L. Ghivelder, A. D. Alvarenga, J. L. Pimentel, Jr., Huiqian Luo, Zhaosheng Wang, and Hai-Hu Wen, Phys. Rev. B {\bf 80}, 014518 (2009).
\bibitem{perkins}G K Perkin, J Moore, Y Bugoslavsky, L F Cohen, J Jun, S M Kazakov, J Karpinski, and A D Caplin, Supercond. Sci. Technol. {\bf 15}, 1156 (2002).
\bibitem{beasley} M. R. Beasley, R. Labash, and W. W. Weeb, Phys. Rev. {\bf 181}, 682 (1969).
\bibitem{donglu} D. Shi and S. Salem-Sugui, Jr., Phys. Rev. B {\bf 44}, 7647 (1991).
\bibitem{tb}S. Salem-Sugui, Jr., A. D. Alvarenga, M. Friesen, K. C. Goretta, O. F. Schilling, F. G. Gandra, B. W. Veal, and P. Paulikas, Phys. Rev. B {\bf 71} 024503 (2005).
\bibitem{tinkham}M. Tinkham, Phys. Rev. Lett. {\bf 61}, 1658 (1988). 
\bibitem{ghiv}S. Salem-Sugui., Jr., L. Ghivelder, M. Friesen, K. Moloni, B. Veal and P. Paulikas, Phys. Rev. B {\bf 60} 102 (1999). 
\bibitem{burlachkov}L. Burlachkov, Phys. Rev. B {\bf 47}, 8056 (1993).

\end{thebibliography}
\end{document}